# Nano-scale tunable optical binding mediated by hyperbolic metamaterials

Natalia A. Kostina[1*], Denis A. Kislov[1], Aliaksandra N. Ivinskaya[1], Alexey Proskurin[1], Dmitrii N. Redka[2], Andrey Novitsky[3], Pavel Ginzburg[4,5,6], and Alexander S. Shalin [1,7*]

[1]Department of Nanophotonics and Metamaterials, ITMO University, Birzhevaja line, 14, 199034 St. Petersburg, Russia

[2]Saint Petersburg Electrotechnical University "LETI" (ETU), 5 Prof. Popova Street, 197376, St. Petersburg, Russia.

[3]Department of Theoretical Physics and Astrophysics, Belarusian State University, Nezavisimosti Avenue 4, 220030 Minsk, Belarus

[4]School of Electrical Engineering, Tel Aviv University, Ramat Aviv, Tel Aviv 69978, Israel

[5]Light-Matter Interaction Centre, Tel Aviv University, Tel Aviv, 69978, Israel

[6]Center for Photonics and 2D Materials, Moscow Institute of Physics and Technology, Dolgoprudny, 141700 Russia

[7]Ulyanovsk State University, Lev Tolstoy Street 42, 432017, Ulyanovsk, Russia

**ABSTRACT:** Carefully designed nanostructures can inspire new type of optomechanical interactions and allow surpassing limitations set by classical diffractive optical elements. Apart from strong near-field localization, nanostructured environment allows controlling scattering channels and might tailor many-body interactions. Here we investigate an effect of optical binding, where several particles demonstrate a collective mechanical behaviour of bunching together in a light field. In contrary to classical binding, where separation distances between particles are diffraction limited, an auxiliary hyperbolic metasurface is shown here to break this barrier by introducing several controllable near-field interaction channels. Strong material dispersion of the hyperbolic metamaterial along with high spatial confinement of optical modes, which it supports, allow achieving superior tuning capabilities and efficient control over binding distances on the nanoscale. In addition, a careful choice of the metamaterial slab's thickness enables decreasing optical binding distances by orders of magnitude compared to free space scenarios due to the multiple reflections of volumetric modes from the substrate. Auxiliary tunable metamaterials, which allow controlling collective optomechanical interactions on the nanoscale, open a venue for new investigations including collective nanofluidic interactions, triggered bio-chemical reactions and many others.

# INTRODUCTION

Optomechanical manipulation [1] is a widely used technique across many disciplines [2,3], where it is utilized for many fundamental and applied investigations. The capability to manipulate small objects with focused light beams and measure pico- and even femto- [4,5] Newton-scale forces opens a venue for studies of new light-matter interaction regimes [6,7] and bio-molecular processes [8,9] to name just few. Conventional optical tweezers realisations rely on diffractive optical elements and, as the result, have limited trapping capabilities in application to nano-scale particles. As a promising paradigm solution, auxiliary nanostructures have been introduced. So-called plasmonic tweezers [10,11], utilize the capability of noble metal structures to confine light beyond the diffraction limit, e.g. [12,13], and provide improved trap stiffness with relatively low optical powers. While majority of plasmonic tweezers configurations utilise nanoantenna arrays, optomechanical surfaces [14,15], metasurfaces [16], and metamaterials [17,18] have been recently proposed. Those types of configurations with less structured features (in comparison to antenna arrays) might provide additional capabilities, such as optical attraction [19], and are less sensitive to accurate positioning of trapping beams in respect to a structure.

In general, functionalities of auxiliary structures can be split into three main categories. The first one is related to the ability of near-field concentration beyond the diffraction limit, which is traced back to the first generation of plasmonic tweezers [20–22]. Here the main tool for analysis is based on dipolar approximation, where the manipulated particle's size is small compared with the fastest spatial intensity variation. It is also important that this model assumes the local field to remain unperturbed by a small particle. The next level of sophistication in auxiliary structures design is to account for a modified density of photonic states, which governs scattering channels from the particle. For example, if a nearby structure significantly modifies a scattering pattern, the particle takes the recoil in order to conserve the entire linear momentum. One of the main functions of metasurfaces and metamaterials [16–18] is to tailor scattering into high density of states modes. Apparently, the most complex approach to optomechanical manipulation utilises active feedback, where a Brownian particle in an optical fields modifies the trapping potential dynamically and experiences a back action effect [23].

An important niche in the field of the opto-mechanical manipulation is devoted to the investigation of many-body interactions mediated by self-consistent optical fields. Light-induced binding of micro- and nanosized objects can provide stable configurations of particles due to light re-scattering and their self-organization under external illumination [24,25]. Capabilities to achieve simultaneous sorting and ordering of particles' clusters without a need to structure the incident beam makes optical binding advantageous over holographic tweezing techniques [3,26,27]. Various binding scenarios have been investigated and include studies of interactions under Gaussian and Bessel shaped beams illumination [28,29], pattern creation with several interfering beams [30,31], evanescent fields excitations, and self-organization of several optically-interacting plasmonic particles [32]. However, those methods rely on either high field intensities or specific particles' materials, which may limit their generality. Increasing optical trap stiffness without a need to use high-intensity illumination, flexible control over interparticle distances as well as anisotropic optical binding in different directions are among the long-standing challenges, valuable from both fundamental and practical standpoints [33]. Parameters of optical binding can be significantly influenced by introducing a nearby interface. It modifies both the incident field due to Fresnel reflection and effective particles' polarizabilities owing to near-field interactions, qualitatively understood with the help of the image theory [20,34,35]. It was shown, that metal-dielectric interfaces supporting the propagation of surface plasmon-polariton modes (SPPs) can increase optical trapping stiffness and reduce particle-interface separation distances owing to strong interactions with SPPs [10,36–38].

Structured interfaces can provide an additional flexibility in tailoring scattering channels via a pre-designed dispersion of surface and bulk modes. Anisotropic response is one among many possibilities. Generally, anisotropic metamaterials [39–41], proved to be useful in various types of applications i.e. cloaking [42–44], super-resolution [45,46], energy transfer [47–49], and recently have opened a venue for flexible optomechanical control. Specifically, it has been shown that hyperbolic dispersion of bulk modes causes optical pulling forces [17], can lead to levitation [50], repulsion [51] and can even generate negative lateral optical forces along the surface [16].

Here we investigate capabilities of hyperbolic metamaterials substrates in application to optical binding. Typical scenario is depicted in Figure 1, where a pair of small particles are linked together by an optical field, mediated by a structured surface. In contrary to free space binding scenarios, layered metal-dielectric substrate opens additional interaction channels, mediated by surface and volumetric modes. As it will be shown, the interplay between the surface geometry and the modes within the bulk, will allow achieving optical binding with deeply subwavelength separation distances and even efficiently tune the latter by exploiting strong chromatic dispersion of the metamaterial.

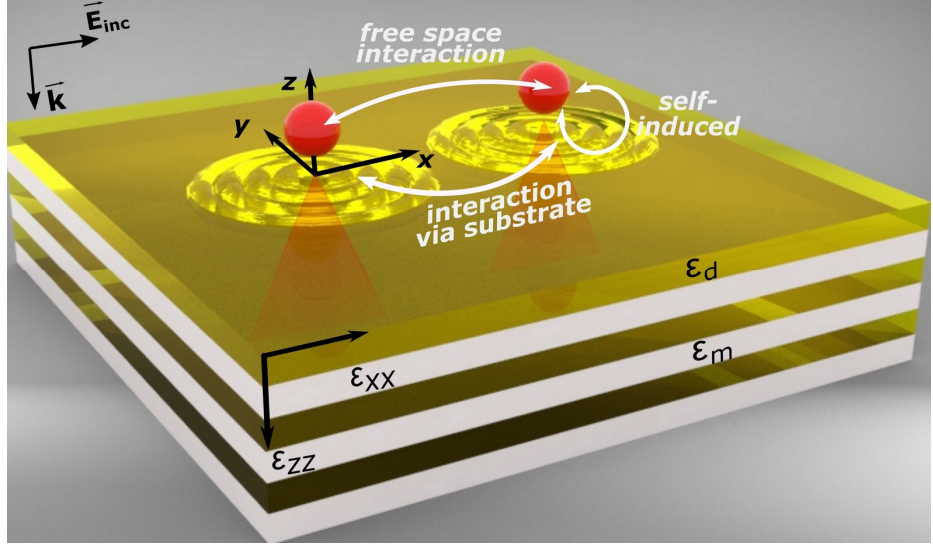

**Figure 1.** The general concept of optical binding above a metamaterial slab. Highly confined optical modes inside the layered hyperbolic metamaterial open additional interaction channels and allow for the formation of dimers and chains with separation distances below the diffraction limit.

The manuscript is organized as follows: Green's function approach to optomechanical interactions is revised first and then followed by the analysis of optical binding next to semi-infinite hyperbolic substrate and finite thickness slab.

GREEN'S FUNCTIONS FORMALISM IN APPLICATION TO OPTICAL BINDING NEAR INTERFACES

The considered scenario is depicted in Figure 1 - a plane wave illuminates two subwavelength nanoparticles placed in a vicinity of an anisotropic substrate. Particles' locations in Cartesian coordinates are $(0,0,a)$ and $(x,y,a)$, where $a = \lambda_0/30$ is the radius of the particles, and $\lambda_0$ is the incident light wavelength. Light-matter interactions with the particles will be analysed under the dipolar approximation. There are three types of channels, which govern the binding phenomenon: (i) particle-particle interaction via the substrate modes, (ii) particle-particle interaction via free space modes, and (iii) individual coupling between each particle and the substrate.

Optical force on a particle in the dipolar approximation can be written as follows:

$$\vec{F} = \frac{1}{2}\operatorname{Re}\sum_i \alpha(\omega)E_i^*(\vec{r},\omega)\nabla E_i(\vec{r},\omega), \qquad (1)$$

where $E_i$ corresponds to the $i^{th}$ component of the self-consistent electric field, $i = x, y, z$ are coordinates, and $\alpha(\omega)$ is the dipolar particle's polarizability in vacuum, including the radiation correction

$$\alpha = \frac{\alpha_{es}}{1 - i\frac{k_0^3}{6\pi\varepsilon_0}\alpha_{es}}; \quad \alpha_{es} = 4\pi\varepsilon_0 a^3 \frac{\varepsilon_p - \varepsilon_1}{\varepsilon_p + 2\varepsilon_1}.$$ $\varepsilon_0$ is the vacuum permittivity, $\varepsilon_p$ is the permittivity of the particle, $\varepsilon_1$ is the permittivity of the surrounding media.

In our notations, where $\vec{r}_1 = (0, 0, a)$ and $\vec{r}_2 = (x, y, a)$, the self-consistent electric field is given by:

$$\vec{E}(\vec{r}_j) = \vec{E}_{inc}(\vec{r}_j) + \frac{k_0^2}{\varepsilon_0}\ddot{G}(\vec{r}_j, \vec{r}_1)\alpha_1 \vec{E}(\vec{r}_1) + \frac{k_0^2}{\varepsilon_0}\ddot{G}(\vec{r}_j, \vec{r}_2)\alpha_2 \vec{E}(\vec{r}_2), \quad j = 1, 2. \quad (2)$$

The first term here represents the incident field with the substrate reflection taken into account, the second and the third terms are the contributions of the dipoles. The Green's function $\ddot{G}$ encapsulates the entire information of the interaction of the dipole with the substrate, e.g., [34]. Substituting the solution of self-consistently formulated Eq. 2 into Eq. 1 allows for calculating the optical force.

TAILORING GREEN'S FUNCTIONS NEAR HYPERBOLIC SUBSTRATES

Investigation of different particle-substrate interaction channels can be performed by analyzing the corresponding Green's function in reciprocal space (k-space). This integral representation, as it will be shown hereinafter, can be split into three parts corresponding to the interaction channels with different physical origin. In particular, propagating (non-evanescent in the upper half-space) modes, surface plasmons, and bulk hyperbolic modes can be involved [16,44,52–54]. Further, we will consider layered realization of the metamaterial depicted in Fig. 1. The permittivity tensor linked to the chosen set of layers is diagonal and obtained via standard homogenization theory [55,56] with $\varepsilon_{xx} = \varepsilon_{yy} \neq \varepsilon_{zz}$ where hyperbolic dispersion occurs when $\text{Re}[\varepsilon_{xx}] < 0$ and $\text{Re}[\varepsilon_{zz}] > 0$. Those components also have strong chromatic dispersion, which will be subsequently used for achieving tunability in binding parameters (see section "Chromatic Tuning of Binding").

In order to split the spectral integral representing the Green's function in the *k*-space, dispersion of the contributing modes should be derived first. Longitudinal component of the wavevector of bulk metamaterial mode has the form of [54]:

$$k_{z2} = \sqrt{(k_0^2 \varepsilon_{zz} - k_x^2)\frac{\varepsilon_{xx}}{\varepsilon_{zz}}}, \quad (3)$$

where $k_0$ is the wave number of an incident wave, $k_x$ is the component of the wavevector of a bulk mode along substrate surface (transversal component). While $\frac{\text{Re}[\varepsilon_{xx}]}{\text{Re}[\varepsilon_{zz}]} < 0$, the wave propagation in a bulk hyperbolic material is possible as long as $k_x$ surpasses a critical value

$$k_{cr} = k_0\sqrt{\varepsilon_{zz}}, \quad (4)$$

and $k_z$ becomes real.

In order to reveal the contribution of different types of modes (free space, plasmons, hyperbolic modes), Fresnel coefficients should be analysed [54]. The reflection coefficient from semi-infinite hyperbolic substrate for s- and p-polarized wave is given by:

$$r^p = \frac{\varepsilon_{xx}k_{z1} - \varepsilon_1 k_{z2}}{\varepsilon_{xx}k_{z1} + \varepsilon_1 k_{z2}}, \quad r^s = \frac{k_{z1} - k_{z2}}{k_{z1} + k_{z2}}, \quad (5)$$

here $\varepsilon_1$ denotes the dielectric permittivity of the upper half-space, $k_{z2}, k_{z1}$ – longitudinal (perpendicular to the substrate) wavevector components in the hyperbolic metamaterial and in the upper half-space respectively.

Examination of the Fresnel coefficients allows identifying conditions for excitation of two types of modes in the structure: volumetric hyperbolic modes in metamaterial and surface plasmon-polariton (SPP) on its interface. From the reflection coefficient for p-polarization it is possible to obtain SPP propagation constant (note, that SPP is naturally p-polarized):

$$k_x^{SPP} = k_0 \sqrt{\frac{\varepsilon_1(\varepsilon_{xx} - \varepsilon_1)}{\varepsilon_{xx} - \frac{\varepsilon_1^2}{\varepsilon_{zz}}}} . \qquad (6)$$

SPP exists only if $\mathrm{Re}[\varepsilon_{zz}] > \mathrm{Re}[\varepsilon_1]$, resulting in imaginary $z$ and real $x$ components of the SPP wavevector. Surface-plasmon polariton resonance condition corresponds to the zero denominator of Eq. (6), but it is not possible as far as $\mathrm{Re}[\varepsilon_{xx}] < 0, \mathrm{Re}[\varepsilon_{zz}] > 0$. The minimal value of the denominator corresponds to the $\mathrm{Re}[\varepsilon_{xx}] \to 0$ and $\mathrm{Re}[\varepsilon_{zz}] \to \infty$, which is close to the surface plasmon-polariton excitation. For the opposite case $\mathrm{Re}[\varepsilon_{xx}] > 0, \mathrm{Re}[\varepsilon_{zz}] < 0$ surface plasmon-polariton does not exist, as perpendicular to the surface wavevector component is real.

Let us consider in details the hyperbolic case of $\mathrm{Re}[\varepsilon_{xx}] < 0, \mathrm{Re}[\varepsilon_{zz}] > 0$. Figure 2 (a) is presented to provide better understanding of the modal structure of the system. Imaginary part of the reflection coefficient contains information about all of the modes [54]. Here the dispersion for homogenized multi-layered Ag/Ta$_2$O$_5$ is presented (filling factor of the structure is 0.133). The imaginary part of the reflection coefficient over the parallel to substrate wavevector component and frequency ω is presented. White lines (solid and dash-dotted) correspond to the light line ($k_0(\omega) = \omega/c$) and critical wavevector $k_{cr}(\omega)$ from Eq. (4), blue line illustrates the dispersion characteristic of the surface plasmon-polariton $k_x^{spp}(\omega)$ from Eq. (6), and hyperbolic modes are marked with white dashed lines (just a few examples). Behaviour of the $\mathrm{Im}[r^p]$ is in a perfect agreement with the dispersion characteristics.

Therefore, there are three important regions governing the interaction of a nanoparticle with a hyperbolic metamaterial. Hyperbolic modes are contributing for $k_x > k_{cr}$, surface plasmon-polaritons are supported between $k_0$ and $k_{cr}$, so the distance between them defines the overall contribution of SPPs, and free-space modes are allowed at $0 \leq k_x \leq k_0$. In the particular case of $k_{cr} \to k_0$ the SPP is negligible and hyperbolic modes and free-space modes play the main role. This scenario (among many others) is considered on Fig. 2 (b) (red line) to underline the contribution of hyperbolic modes at the absence of SPPs.

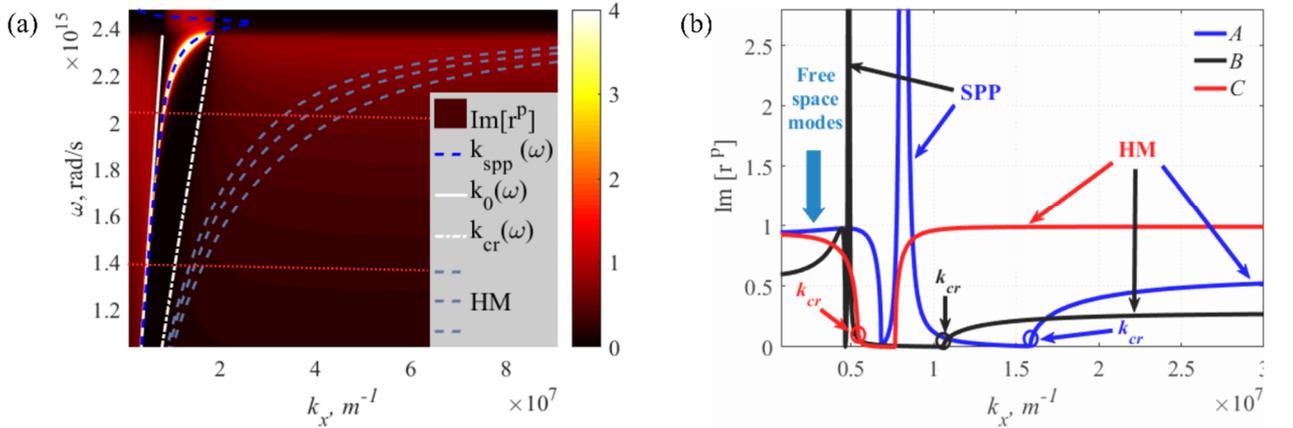

**Figure 2.** (a): colour map of the imaginary part of the reflection coefficient over wavevector's x-component and incident wave frequency. From this graph one can, e. g., pick out frequencies $\omega = 2.05 \cdot 10^{15}$ and $\omega = 1.4 \cdot 10^{15}$ rad/s as points A and B correspondingly (shown with dotted red lines). A corresponds to $\lambda_0 = 920$ nm and effective medium parameters $\varepsilon_{xx} = -1.714 + 0.075i$, $\varepsilon_{zz} = 5.392 + 0.0084i$ *and for* B $\lambda_0 = 1350$ nm, $\varepsilon_{xx} = -8.94 + 0.33i$, $\varepsilon_{zz} = 5.19 + 0.0118i$. (b): Imaginary part of the reflection coefficient as the function of the wavevector x-component. The dependence is plotted for three sets of parameters: A with blue, B with black and C (ideal case without the SPPs contribution for $\lambda_0 = 920$ nm, $\varepsilon_{xx} = -2 + 0.066i$, $\varepsilon_{zz} = 0.5 + 0.0084i$) with red lines. Characteristic regions to underline the contributions of different interaction channels are: $k_x \in [0; k_0]$ for propagating free-space modes, $k_x \in (k_0; k_{cr}]$ for SPP, $k_x \in (k_{cr}, \infty)$ for hyperbolic modes.

Thus, Green's function for particles-substrate interaction can be decomposed as follows:

$$\vec{G}^{subs}(\vec{r}_i, \vec{r}_j) = \int_0^{k_0} \vec{M}^{subs}(k_x) dk_x + \int_{k_0}^{k_{cr}} \vec{M}^{subs}(k_x) dk_x + \int_{k_{cr}}^{\infty} \vec{M}^{subs}(k_x) dk_x, \quad j=1,2; i=1,2. \quad (7)$$

The integrand matrix $\vec{M}^{subs}$ in the Green's function is presented in Supplementary Material (Section 1).

In accordance with the aforementioned: $I = \int_0^{k_0}$ - the free-space propagating modes contribution, $II = \int_{k_0}^{k_{cr}}$ - the surface plasmon-polariton contribution (if SPPs are supported $\varepsilon_{zz} > \varepsilon_1$), and $III = \int_{k_{cr}}^{\infty}$ - volumetric (hyperbolic) modes of the substrate. The hyperbolic modes contribution is usually estimated with the approximation $k_x / k_0 \to \infty$, where reflection from a substrate depends only on the dielectric permittivities, as long as $k_x > k_0$ [16,50,51]. Moreover, the interplay between plasmonic and hyperbolic contributions could be efficiently tailored via adjusting material parameters ($\varepsilon_{zz}$) and, consequently, $k_{cr}$.

In order to demonstrate this capability, the imaginary part of the reflection coefficient for p-polarized wave as the function of $k_x$ has been plotted in Fig. 2 (b) for different ω corresponding to different material parameters: Line A ($\lambda_0 = 920$ nm, $\varepsilon_{xx} = -1.714 + 0.075i$, $\varepsilon_{zz} = 5.392 + 0.0084i$), where $k_{cr}$ is quite large and the SPP contribution is dominating, Line B ($\lambda_0 = 920$ nm, $\varepsilon_{xx} = -1.714 + 0.075i$, $\varepsilon_{zz} = 5.392 + 0.0084i$), where SPPs peak is much narrower, and bulk modes contribution is more pronounced, and Line C for arbitrary metamaterial with ($\lambda_0 = 920$ nm, $\varepsilon_{xx} = -2 + 0.066i$, $\varepsilon_{zz} = 0.5 + 0.0084i$), where $k_{cr}$ is less then $k_0$, the SPP contribution is absent, and, consequently, the interaction is governed by free space modes and bulk hyperbolic modes (magnitudes are related as $III = 2.5\ I$). These particular scenarios will be further investigated in terms of optical forces. We should stress, that each of these integrals is taken into account twice via effective field Eq. (1), so the overall difference in force is bigger.

Noteworthy, the interval-based integration given by Eq. (7) stays clear only for a standalone particle. Introducing another one involves cross-coupling between different terms, e.g., SPP generated by the first particle could be scattered by another one into bulk hyperbolic modes and vice versa. This effect will be considered in the next section and shown to have minor impact on the overall trapping and binding efficiency.

RESULTS

Having identified the contribution of different interaction channels to the Green's function, we can proceed with the self-consistent scattering problem (Eq. (2)).

Semi-infinite substrate

Influence of a semi-infinite anisotropic multi-layered metamaterial on optical binding will be analysed next. The most significant parameters for binding are the period and stiffness allowing for effective structuring of nanoparticles in many different 2D and even 3D architectures [25,31,57]. Recently, we revealed the possibility to bind nanoparticles with subwavelength separation distances via the interference of surface plasmon-polaritons [34]. Introducing additional metamaterial bulk modes seems promising for the further enhancement of binding.

Let us consider a pair of nanoparticles, one of which is fixed at the origin of the coordinates as in the previous scenario. For the sake of simplicity, we consider the second particle to have the same parameters as the first one. Effective field at the nanoparticle follows from Eq. (2) and is given in Supplementary (Section 2).

The period of optical binding can now be defined as a distance between two nearest stable equilibrium positions, and the stiffness is the ratio of the restoring force to the particle's displacement $\kappa = -\Delta F_x / \Delta x$ (in a close vicinity of a stable position, where $F_x(x)$ is close to a linear profile). Hereinafter the period $L_{bind}$ and distances will be normalized over the incident wavelength $\lambda_0$, and the optical forces - over the radiation pressure $F_0 = \frac{\text{Im}(\alpha)}{2} k_0 |\vec{E}_{inc}|^2$.

The material parameters are taken to be the same as for lines *A, C* in Fig. 2 (b) corresponding to the dominating contributions of SPPs (*A*), and hyperbolic modes (*C*). In Fig. 3 the optical forces for both principally different scenarios are shown. The blue lines correspond to the total optical force, the black lines correspond to the contribution of modes with $k_x \leq k_0$ (propagating free-space modes in the upper half-space). The SPP contribution for *A* is given by the red line (it is zero for *C* case by definition, see the previous section). HM contribution is depicted by grey circles.

In the case *A* governed by the strong impact of surface plasmon-polaritons the optical forces are fully driven by these surface waves, and the contribution of other modes is insufficient. Moreover, in the case *C* with the predominating influence of the hyperbolic modes, optical binding has almost nothing special in comparison with the free-space scenario. In this case, HMs contribution just increases the force almost twice (which is still 2 orders of magnitude less than that of SPPs) and slightly shifts equilibrium positions almost not affecting $L_{bind}$. In this case of a semi-infinite metamaterial for normally incident light the hyperbolic modes excited by the first particle just propagate symmetrically in the volume not interacting with the second particle, and vice versa.

However, the still existing nonzero contribution of HMs can be explained via the aforementioned cross-terms, when modes excited by one particle are scattered by another one giving rise to additional HMs with broken symmetry, which, in turn, lead to the optical force shift [16].

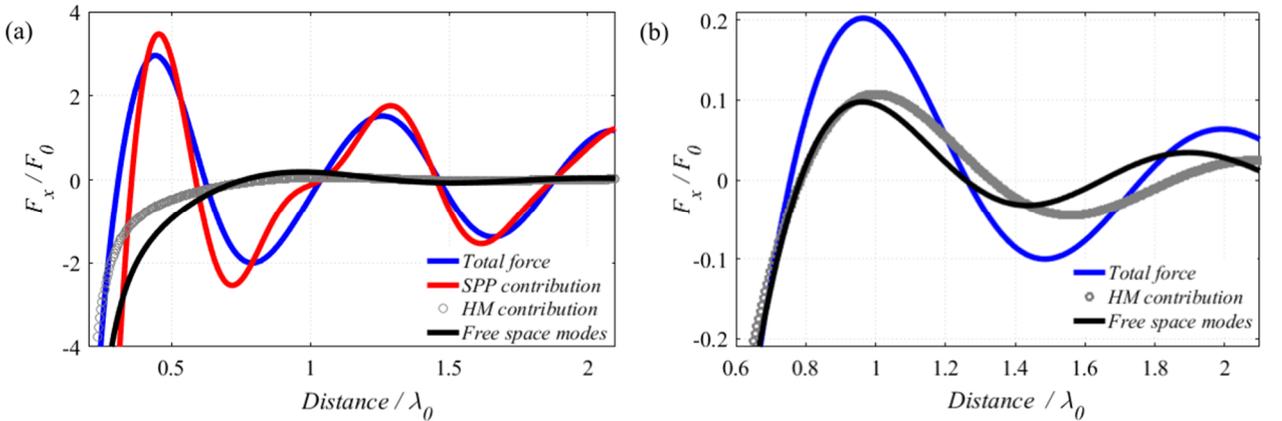

**Figure 3.** Dependence of the optical binding force on the distance between the particles. (a) is for parameters A from Fig. 2 (b). (b) is for line C. The blue line corresponds to the total optical binding force near anisotropic substrate, the red line is for the surface plasmon-polariton modes contribution only, grey circles depict contribution of hyperbolic modes, and the black lines show optical binding via propagating in the upper half-space modes only.

Thus, the hyperbolic modes even being dominating in the interaction with the semi-infinite metamaterial do not provide a sufficient contribution to binding in this case.

Finite thickness metamaterial slabs

The main reason for the weak influence of HMs on binding is the lack of a feedback from the bulk modes, which propagate away from the particles to infinity. However, as it will be shown hereinafter, strong optical binding can be obtained utilizing anisotropic *finite thickness* slab due to reflections of hyperbolic modes from the boundaries. The structure under consideration appears in Figure 4.

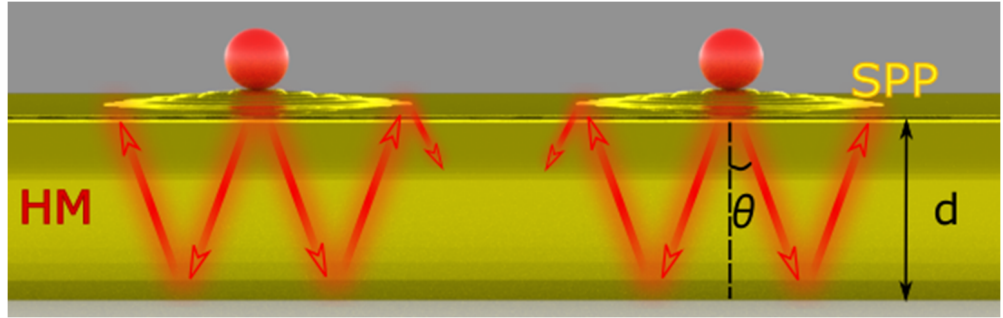

**Figure 4.** The scheme of optical binding near anisotropic hyperbolic metamaterial (HM) slab. Reflections from the boundaries of the slab form high intensity regions and result in optical binding with separation distances $L_{bind}$ below the diffraction limit.

In contrast to conventional waveguides, $k_x$ in hyperbolic slabs can achieve rather high values, which, together with highly confined shape of the modes, could allow for very small pdistances beween the hot-spots driven by multiple reflections. This, in turn, paves a way for strongly subwavelength binding of nanopartiles and also provides tunability via changing material parameters, slab thickness, excitation wavelength etc.

The formalism developed for the semi-infinite substrate is also applicable for the finite-thickness slab. The main difference is in the Fresnel reflection coefficient, which in the latter case is given by:

$$r_{slab} = \frac{r - r\exp(2ik_{2z}d)}{1 - (r)^2 \exp(2ik_{2z}d)} \quad (8)$$

where $d$ represents the thickness of the slab. It is clearly seen, that additional periodical maxima will be present in $r^p_{slab}(k_x)$ (reflection of p-polarised wave). These peaks are related to the additional boundary, which causes multiple reflections between upper and lower interfaces. The distance between hot spots at the interface, and, thus, between the bound particles depends on the parameter $d$ and angle between group velocity of the hyperbolic modes and the normal to the surface. Noteworthy, that hyperbolic modes are not usual "geometric" beams, thus exact calculations are needed to find actual binding period $L_{bind}$.

Let us consider optical binding force near anisotropic slab with parameters $A$, $C$ (from Fig. 2 (b)) and thicknesses $d = \lambda_0/2$, $d = \lambda_0/8$, $\lambda_0 = 920$ nm. Figure 5 represents imaginary part of the reflection coefficient (left column, (a, c)), and optical force (right column, (b, d)). Noteworthy, additional peaks corresponding to the multiple reflections appear in the reflection coefficient. Here we show $\text{Im}\left[r^p\right]$ only for $k_x/k_0 \leq 10$, because the next peaks are much weaker due to the absorption and are not necessary for subsequent qualitative analysis. However, the force calculations take into account all possible $k_x$: $0 \leq k_x/k_0 < \infty$ to provide accurate values. The distance between the reflection peaks increases (in *k*-space) with decreasing of the thickness and leads to the optical force period decrease.

Comparing Figs 2 (b) and 5 (a) we find SPP contribution to become much less pronounced (magnitudes of integrals from formula (7) $II \approx 3\,I$ for semi-infinite case, $II \approx 2.3\,I$ for $d = \lambda_0/2$ and $II \approx 2\,I$ for $d = \lambda_0/8$ (Fig. 5 (a) )) (where $I$ is the integral contribution of the free-space modes from Eq. 7). For thin slab $d = \lambda_0/8$ the contribution of SPPs $k_x/k_0 = 1.044$ and $k_x/k_0 = 1.84$ can be considered as negligible for small distances and almost does not influence the optical forces (Fig. 5 (b)) governed predominantly by the hyperbolic mode with $k_x/k_0 = 8.8$ ( $III \approx 7\,I$ ) and $L_{bind} \approx 1/8.8 \approx 0.114$. For $d = \lambda_0/2$, however, we have more reach modes kit, e.g., SPPs ($k_x/k_0 = 1.17$ and $k_x/k_0 = 1.2$) together with a set of hyperbolic modes. This leads to the peculiar behaviour of the optical forces: HM with $k_x/k_0 \geq 3.08$ enables subwavelenght binding with $L_{bind} \approx 1/3.08 \approx 0.32$ modulated by SPPs overall envelope (schematically shown with gray dotted line) with the period $\approx 0.85$. Note, all the distances are normalized over the wavelength of the incident wave $\lambda_0$.

For the parameters $C$ (Fig. 5 (c, d)) there is no surface plasmon-polariton contribution by definition (we have chosen the appropriate parameters in the first section especially to emphasize the influence of hyperbolic modes in both semi-infinite and finite cases), because $k_{cr}/k_0 = 0.7 < 1$, so the optical forces are completely dependent on the free-space and volumetric modes. For the half-wavelength hyperbolic slab the forces are governed by free-space propagating waves ( $II = 0$, $III \approx I$ ) with small $k_x$, and low amplitude HMs (high $k_x$ ones are effectively absorbed via rather large thickness), thus, the dependence is close to optical binding in a free space both in terms of period and forces magnitude. For the thin slab the force is almost 2 orders of magnitude increased (Fig. 5 (d)) and the periodicity now is fully driven by the two most pronounced peaks $k_x/k_0 = 1.33$ ( $L_{bind} \approx 0.75$ ) shown by gray envelope, and $k_x/k_0 = 3.11$ corresponding to $L_{bind} \approx 0.32$ (overall integral $III \approx 6.3\,I$ ).

The aforementioned qualitative mode analysis of optical forces and binding period is an approximation. Often, several peaks contribute to the optical force and form a unique signature, either with SPP or not (e.g., curved and asymmetric Figs 5(b, d)). However, it allows for better understanding of the binding scenarios at the presence of such complicated structures as hyperbolic metamaterials and even for some quantitative estimations provided above.

Moreover, comparing two principally different scenarios (with and without SPPs' contributions), we find a new possibility to obtain optical binding forces, which are several orders of magnitude higher than in the free space binding scenarios (and about 1 order of magnitude stronger than that delivered by plasmonic metals [34]). Furthermore, tuning the distance between bound particles beyond the diffraction limit is also possible. These characteristics are strongly enhanced in the case of thin slabs allowing for better utilizing hyperbolic modes feedback, and the proved great tolerance to the metamaterial parameters paves a way to a plethora of highly demanded application.

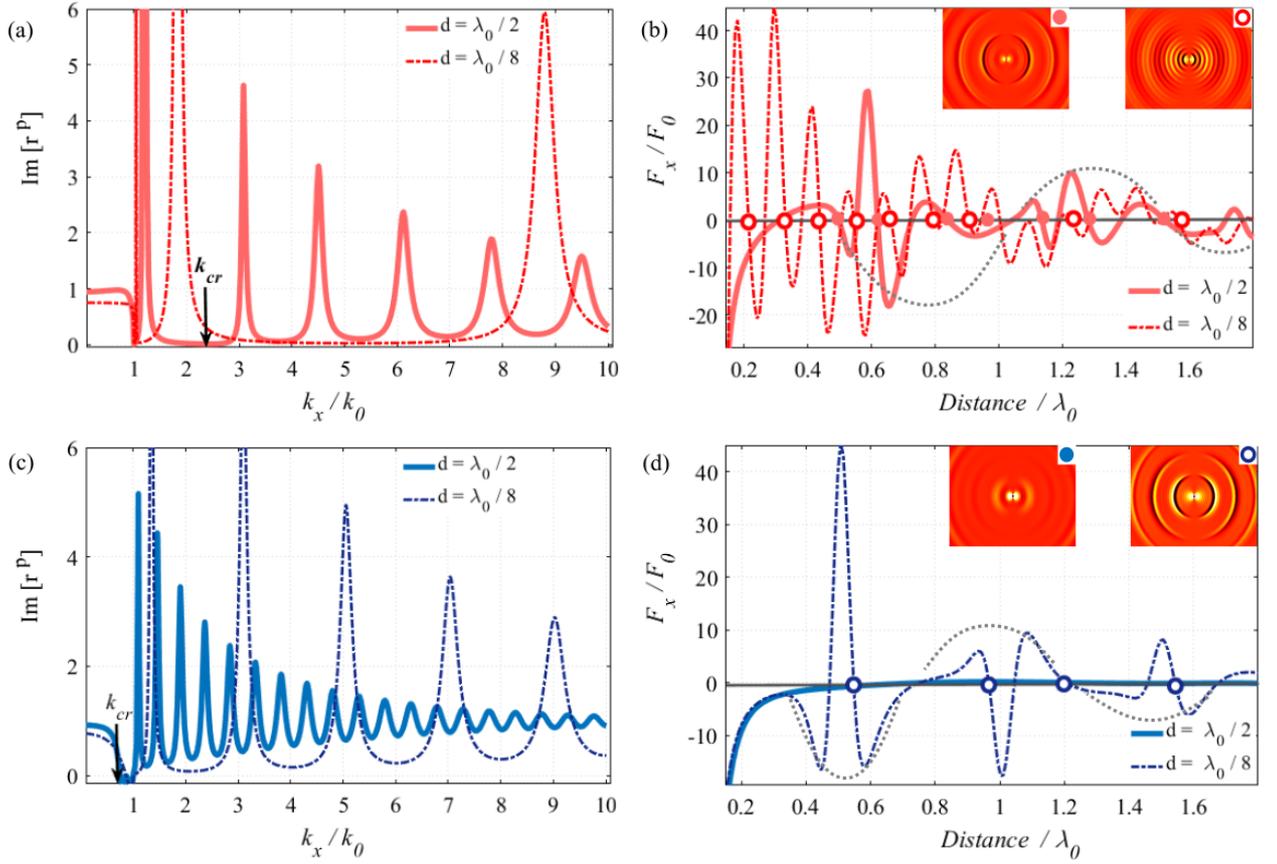

**Figure 5.** Imaginary part of the reflection coefficient over relative transversal wavevector component (a, c) and optical binding force (b, d). The dependence is plotted for two sets of parameters: (a, b) correspond to the parameters A from Fig. 2. (c, d) correspond to C from Fig. 2. The solid lines depict reflection coefficient and optical binding force for slab with thickness $d = \lambda_0/2$, the dash dotted lines depict the reflection coefficient for slab with the thickness $d = \lambda_0/8$, where $\lambda_0 = 920$ nm is the incident wavelength. The black arrows in the (a, c) show position of the $k_{cr}$. One can see that for C it is placed before $k_0$. Gray dotted lines represent envelopes (see text). The insets in (b, d) show electric component $E_x$ of the field scattered by the particle above the homogeneous hyperbolic substrate with thickness $d = \lambda_0/2$ (left column) and $d = \lambda_0/8$ (right column).

Chromatic tuning of binding

In the previous sections we have considered semi-infinite and finite slabs of hyperbolic metamaterials consisting of the Ag [58] and $Ta_2O_5$ [59] layers allowing for effective tuning via adjusting material and geometrical parameters. Hereinafter let us consider another important degree of freedom – chromatic tuning of the metamaterial-assisted optical binding.

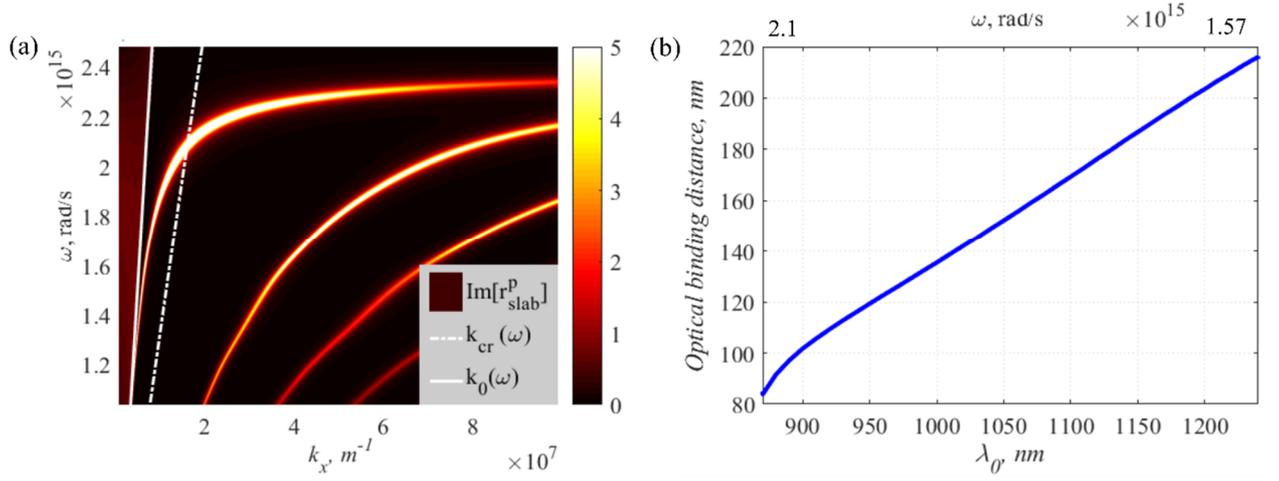

**Figure 6.** Chromatic tuning for the multilayered structure of silver and $Ta_2O_5$ [58,59] layers with the slab thickness 115 nm and filling factor of 0.133. (a) - imaginary part of the reflection coefficient (dispersion diagram). (b) - optical binding period over the frequency.

The figure 6 (a) shows the dependence of the reflection coefficient on the incident wave frequency and $k_x/k_0$ for the slab thickness 115 nm. It can be seen, that the number of HM peaks governed by the reflections (equally to Fabry-Perot resonances for hyperbolic modes) and contributing to optical binding is increased with lower frequency ($d/\lambda_0$ decrease), so the optical force dependence becomes more complicated.

The distance between the bound particles (Fig. 6 (b)) is now a function of the frequency, thus the materials dispersion plays a key role here. The binding period is proportional to the relation of the thickness of the slab and incident wavelength and $\varepsilon_{xx}(\omega)$, $\varepsilon_{zz}(\omega)$. In this case the permittivities are monotonically dependent on the frequency [39], thus the dependence of the optical binding distance is more or less monotonous. However, in other wavelength regions additional HMs and non-monotonous dispersion of optical constants could displace the stable equilibrium positions and change the dependence shown in Fig. 6 (b). This additional degree of freedom opens a room of opportunities for tuning optical binding via "non-invasive" way and fabricate novel designs and architectures of nanostructures on metamaterial substrates by adjusting optically induced forces with hyperbolic modes.

Moreover, in Supplementary (Section 3) we consider the dependence of the optical binding force on the topmost layer of the slab (metal or dielectric), which also can be useful in a plethora of applications.

CONCLUSION

In this work we describe the transverse optical binding of two particles near hyperbolic metamaterial. High-$k$ volumetric modes can provide additional channels of the particles' interaction with substrates and, therefore, drastically enhance capabilities of optomechanical manipulation schemes. For semi-infinite (or rather thick) metamaterial slab the hyperbolic modes even being dominating in scattering do not contribute to optical binding because of the almost absent feedback (hyperbolic modes excited by one particle do not interact with the second one). In contrast, thin metamaterial slabs provide multiple reflections from boundaries forming a set of strongly localized hot spots with huge intensity gradients governing nanoparticles motion at nanoscale. Furthermore, mode analysis shows the predominant impact of HMs on binding giving rise to several orders of magnitude increased optical forces and deeply subwavelength nanoparticles positioning. Moreover, the principal realization of this phenomenon appears to be rather tolerant to the metamaterial parameters, enabling strongly enhanced performance for a whole set of designs and driving broadband chromatic tuning. Novel auxiliary carefully designed metamaterials and metasurfaces featuring superior optomechanical mechanisms are nowadays extremely demanded in a variety of applications, such as microfluidics, lab-on-chip devices, and biology and medicine to name just few.

Funding Sources

The work has been supported in part by ERC StG 'In Motion' and PAZY Foundation. A.S.S acknowledges the support of the Russian Fund for Basic Research within the projects 18-02-00414, 18-52-00005. Simulations of the optical forces have been supported by the Russian Science Foundation (Project No. 18-72-10127). AN thanks the Belarusian Republican Foundation for Fundamental Research (Project No. F18R-021).

**REFERENCES**:

(1) Ashkin, A. Acceleration and Trapping of Particles by Radiation Pressure. *Phys. Rev. Lett.* **1970**, *24* (4), 156–159. https://doi.org/10.1103/PhysRevLett.24.156.

(2) Neuman, K. C.; Block, S. M. Optical Trapping. *Rev. Sci. Instrum.* **2004**, *75* (9), 2787–2809. https://doi.org/10.1063/1.1785844.

(3) MacDonald, M. P.; Paterson, L.; Volke-Sepulveda, K.; Arlt, J.; Sibbett, W.; Dholakia, K. Creation and Manipulation of Three-Dimensional Optically Trapped Structures. *Science.* **2002**, *296* (5570), 1101–1103. https://doi.org/10.1126/science.1069571.

(4) Jonáš, A.; Zemánek, P. Light at Work: The Use of Optical Forces for Particle Manipulation, Sorting, and Analysis. *Electrophoresis* **2008**, *29* (24), 4813–4851. https://doi.org/10.1002/elps.200800484.

(5) Damková, J.; Chvátal, L.; Ježek, J.; Oulehla, J.; Brzobohatý, O.; Zemánek, P. Enhancement of the "tractor-Beam" Pulling Force on an Optically Bound Structure. *Light Sci. Appl.* **2018**, *7* (1), 17135. https://doi.org/10.1038/lsa.2017.135.

(6) Novotny, L.; Bian, R. X.; Xie, X. S. Theory of Nanometric Optical Tweezers. *Phys. Rev. Lett.* **1997**, *79* (4), 645–648. https://doi.org/10.1103/PhysRevLett.79.645.

(7) Juan, M. L.; Gordon, R.; Pang, Y.; Eftekhari, F.; Quidant, R. Self-Induced Back-Action Optical Trapping of Dielectric Nanoparticles. *Nat. Phys.* **2009**, *5* (12), 915–919. https://doi.org/10.1038/nphys1422.

(8) Grier, D. G. A Revolution in Optical Manipulation. *Nature* **2003**, *424* (6950), 810–816. https://doi.org/10.1038/nature01935.

(9) Moffitt, J. R.; Chemla, Y. R.; Smith, S. B.; Bustamante, C. Recent Advances in Optical Tweezers. *Annu. Rev. Biochem.* **2008**, *77* (1), 205–228. https://doi.org/10.1146/annurev.biochem.77.043007.090225.

(10) Juan, M. L.; Righini, M.; Quidant, R. Plasmon Nano-Optical Tweezers. *Nat. Photonics* **2011**, *5* (6), 349–356. https://doi.org/10.1038/nphoton.2011.56.

(11) Righini, M.; Volpe, G.; Girard, C.; Petrov, D.; Quidant, R. Surface Plasmon Optical Tweezers: Tunable Optical Manipulation in the Femtonewton Range. *Phys. Rev. Lett.* **2008**, *100* (18), 186804. https://doi.org/10.1103/PhysRevLett.100.186804.

(12) Gramotnev, D. K.; Bozhevolnyi, S. I. Plasmonics beyond the Diffraction Limit. *Nat. Photonics* **2010**, *4* (2), 83–91. https://doi.org/10.1038/nphoton.2009.282.

(13) Berkovitch, N.; Ginzburg, P.; Orenstein, M. Nano-Plasmonic Antennas in the near Infrared Regime. *J. Phys. Condens. Matter* **2012**, *24* (7), 073202. https://doi.org/10.1088/0953-8984/24/7/073202.

(14) Shilkin, D. A.; Lyubin, E. V.; Soboleva, I. V.; Fedyanin, A. A. Trap Position Control in the Vicinity of Reflecting Surfaces in Optical Tweezers. *JETP Lett.* **2014**, *98* (10), 644–647. https://doi.org/10.1134/S0021364013230124.

(15) Ivinskaya, A.; Petrov, M. I.; Bogdanov, A. A.; Shishkin, I.; Ginzburg, P.; Shalin, A. S. Plasmon-Assisted Optical Trapping and Anti-Trapping. *Light Sci. Appl.* **2017**, *6* (5), e16258. https://doi.org/10.1038/lsa.2016.258.

(16) Ivinskaya, A.; Kostina, N.; Proskurin, A.; Petrov, M. I.; Bogdanov, A. A.; Sukhov, S.; Krasavin, A. V.; Karabchevsky, A.; Shalin, A. S.; Ginzburg, P. Optomechanical Manipulation with Hyperbolic Metasurfaces. *ACS Photonics* **2018**, *5* (11), 4371–4377. https://doi.org/10.1021/acsphotonics.8b00775.

(17) Bogdanov, A. A.; Shalin, A. S.; Ginzburg, P. Optical Forces in Nanorod Metamaterial. *Sci. Rep.* **2015**, *5*, 15846. https://doi.org/10.1038/srep15846.

(18) Shalin, A. S.; Sukhov, S. V.; Bogdanov, A. A.; Belov, P. A.; Ginzburg, P. Optical Pulling Forces in Hyperbolic Metamaterials. *Phys. Rev. A - At. Mol. Opt. Phys.* **2015**, *91* (6), 063830. https://doi.org/10.1103/PhysRevA.91.063830.

(19) Chen, J.; Ng, J.; Lin, Z.; Chan, C. T. Optical Pulling Force. *Nat. Photonics* **2011**, *5* (9), 531–534. https://doi.org/10.1038/nphoton.2011.153.

(20) Shalin, A. S.; Sukhov, S. V. Plasmonic Nanostructures as Accelerators for Nanoparticles: Optical Nanocannon. *Plasmonics* **2013**, *8* (2), 625–629. https://doi.org/10.1007/s11468-012-9447-0.

(21) Shalin, A. S.; Ginzburg, P.; Belov, P. A.; Kivshar, Y. S.; Zayats, A. V. Nano-Opto-Mechanical Effects in Plasmonic Waveguides. *Laser Photonics Rev.* **2014**, *8* (1), 131–136. https://doi.org/10.1002/lpor.201300109.

(22) Baryshnikova, K. V.; Novitsky, A.; Evlyukhin, A. B.; Shalin, A. S. Magnetic Field Concentration with Coaxial Silicon Nanocylinders in the Optical Spectral Range. *J. Opt. Soc. Am. B* **2017**, *34* (7), D36. https://doi.org/10.1364/josab.34.000d36.

(23) Descharmes, N.; Dharanipathy, U. P.; Diao, Z.; Tonin, M.; Houdré, R. Observation of Backaction and Self-Induced Trapping in a Planar Hollow Photonic Crystal Cavity. *Phys. Rev. Lett.* **2013**, *110* (12), 123601. https://doi.org/10.1103/PhysRevLett.110.123601.

(24) Burns, M. M.; Fournier, J. M.; Golovchenko, J. A. Optical Binding. *Phys. Rev. Lett.* **1989**, *63* (12), 1233–1236. https://doi.org/10.1103/PhysRevLett.63.1233.

(25) Dholakia, K.; Zemánek, P. Colloquium: Gripped by Light: Optical Binding. *Rev. Mod. Phys.* **2010**, *82* (2), 1767–1791. https://doi.org/10.1103/RevModPhys.82.1767.

(26) Sukhov, S.; Shalin, A.; Haefner, D.; Dogariu, A. Actio et Reactio in Optical Binding. *Opt. Express* **2015**, *23* (1), 247. https://doi.org/10.1364/oe.23.000247.

(27) Grzegorczyk, T. M.; Kemp, B. A.; Kong, J. A. Stable Optical Trapping Based on Optical Binding Forces. *Phys. Rev. Lett.* **2006**, *96* (11), 1–4. https://doi.org/10.1103/PhysRevLett.96.113903.

(28) Taylor, J. M.; Love, G. D. Optical Binding Mechanisms: A Conceptual Model for Gaussian Beam Traps. *Opt. Express* **2009**, *17* (17), 15381. https://doi.org/10.1364/oe.17.015381.

(29) Chen, J.; Ng, J.; Wang, P.; Lin, Z. Analytical Partial Wave Expansion of Vector Bessel Beam and Its Application to Optical Binding: Erratum. *Opt. Lett.* **2011**, *36* (7), 1243. https://doi.org/10.1364/ol.36.001243.

(30) Simpson, S. H.; Zemánek, P.; Maragò, O. M.; Jones, P. H.; Hanna, S. Optical Binding of Nanowires. *Nano Lett.* **2017**, *17* (6), 3485–3492. https://doi.org/10.1021/acs.nanolett.7b00494.

(31) Donato, M. G.; Brzobohatý, O.; Simpson, S. H.; Irrera, A.; Leonardi, A. A.; Lo Faro, M. J.; Svak, V.; Maragò, O. M.; Zemánek, P. Optical Trapping, Optical Binding, and Rotational Dynamics of Silicon Nanowires in Counter-Propagating Beams. *Nano Lett.* **2019**, *19* (1), 342–352. https://doi.org/10.1021/acs.nanolett.8b03978.

(32) Demergis, V.; Florin, E. L. Ultrastrong Optical Binding of Metallic Nanoparticles. *Nano Lett.* **2012**, *12* (11), 5756–5760. https://doi.org/10.1021/nl303035p.

(33) Taylor, J. M. *Optical Binding Phenomena: Observations and Mechanisms*; 2011. https://doi.org/10.1007/978-3-642-21195-9.

(34) Kostina, N.; Petrov, M.; Ivinskaya, A.; Sukhov, S.; Bogdanov, A.; Toftul, I.; Nieto-Vesperinas, M.; Ginzburg, P.; Shalin, A. Optical Binding via Surface Plasmon Polariton Interference. *Phys. Rev. B* **2019**, *99* (12), 125416. https://doi.org/10.1103/PhysRevB.99.125416.

(35) Chaumet, P. C.; Nieto-Vesperinas, M. Optical Binding of Particles with or without the Presence of a Flat Dielectric Surface. *Phys. Rev. B - Condens. Matter Mater. Phys.* **2001**, *64* (3), 035422. https://doi.org/10.1103/PhysRevB.64.035422.

(36) Quidant, R. Plasmonic Tweezers-The Strength of Surface Plasmons. *MRS Bull.* **2012**, *37* (8), 739–744. https://doi.org/10.1557/mrs.2012.172.

(37) Song, Y. G.; Han, B. M.; Chang, S. Force of Surface Plasmon-Coupled Evanescent Fields on Mie Particles. *Opt. Commun.* **2001**, *198* (1–3), 7–19. https://doi.org/10.1016/S0030-4018(01)01484-5.

(38) Quidant, R.; Girard, C. Surface-Plasmon-Based Optical Manipulation. *Laser Photonics Rev.* **2008**, *2* (1–2), 47–57. https://doi.org/10.1002/lpor.200710038.

(39) Shekhar, P.; Atkinson, J.; Jacob, Z. Hyperbolic Metamaterials: Fundamentals and Applications. *Nano Converg.* **2014**, *1* (1), 14. https://doi.org/10.1186/s40580-014-0014-6.

(40) Ginzburg, P.; Fortuño, F. J. R.; Wurtz, G. A.; Dickson, W.; Murphy, A.; Morgan, F.; Pollard, R. J.; Iorsh, I.; Atrashchenko, A.; Belov, P. A.; et al. Manipulating Polarization of Light with Ultrathin Epsilon-near-Zero Metamaterials. *Opt. Express* **2013**, *21* (12), 14907. https://doi.org/10.1364/oe.21.014907.

(41) Ferrari, L.; Wu, C.; Lepage, D.; Zhang, X.; Liu, Z. Hyperbolic Metamaterials and Their Applications. *Prog. Quantum Electron.* **2015**, *40*, 1–40. https://doi.org/10.1016/j.pquantelec.2014.10.001.

(42) Milton, G. W.; Nicorovici, N. A. P. On the Cloaking Effects Associated with Anomalous Localized Resonance. *Proc. R. Soc. A Math. Phys. Eng. Sci.* **2006**, *462* (2074), 3027–3059. https://doi.org/10.1098/rspa.2006.1715.

(43) Schurig, D.; Mock, J. J.; Justice, B. J.; Cummer, S. A.; Pendry, J. B.; Starr, A. F.; Smith, D. R. Metamaterial Electromagnetic Cloak at Microwave Frequencies. *Science.* **2006**, *314* (5801), 977–980. https://doi.org/10.1126/science.1133628.

(44) Shalin, A. S.; Ginzburg, P.; Orlov, A. A.; Iorsh, I.; Belov, P. A.; Kivshar, Y. S.; Zayats, A. V. Scattering Suppression from Arbitrary Objects in Spatially Dispersive Layered Metamaterials. *Phys. Rev. B - Condens. Matter Mater. Phys.* **2015**, *91* (12), 125426. https://doi.org/10.1103/PhysRevB.91.125426.

(45) Liu, Z.; Lee, H.; Xiong, Y.; Sun, C.; Zhang, X. Far-Field Optical Hyperlens Magnifying Sub-Diffraction-Limited Objects. *Science.* **2007**, *315* (5819), 1686. https://doi.org/10.1126/science.1137368.

(46) Rho, J.; Ye, Z.; Xiong, Y.; Yin, X.; Liu, Z.; Choi, H.; Bartal, G.; Zhang, X. Spherical Hyperlens for Two-Dimensional Sub-Diffractional Imaging at Visible Frequencies. *Nat. Commun.* **2010**, *1* (9), 143. https://doi.org/10.1038/ncomms1148.

(47) Tumkur, T. U.; Kitur, J. K.; Bonner, C. E.; Poddubny, A. N.; Narimanov, E. E.; Noginov, M. A. Control of Förster Energy Transfer in the Vicinity of Metallic Surfaces and Hyperbolic Metamaterials. *Faraday Discuss.* **2015**, *178*, 395–412. https://doi.org/10.1039/C4FD00184B.

(48) Biehs, S. A.; Tschikin, M.; Ben-Abdallah, P. Hyperbolic Metamaterials as an Analog of a Blackbody in the near Field. *Phys. Rev. Lett.* **2012**, *109* (10), 104301. https://doi.org/10.1103/PhysRevLett.109.104301.

(49) Roth, D. J.; Nasir, M. E.; Ginzburg, P.; Wang, P.; Le Marois, A.; Suhling, K.; Richards, D.; Zayats, A. V. Förster Resonance Energy Transfer inside Hyperbolic Metamaterials. *ACS Photonics* **2018**, *5* (11), 4594–4603. https://doi.org/10.1021/acsphotonics.8b01083.

(50) Rodríguez-Fortuño, F. J.; Vakil, A.; Engheta, N. Electric Levitation Using Ïμ -near-Zero Metamaterials. *Phys. Rev. Lett.* **2014**, *112* (3), 033902. https://doi.org/10.1103/PhysRevLett.112.033902.

(51) Rodríguez-Fortuno, F. J.; Zayats, A. V. Repulsion of Polarised Particles from Anisotropic Materials with a Near-Zero Permittivity Component. *Light Sci. Appl.* **2016**, *5* (1), e16022. https://doi.org/10.1038/lsa.2016.22.

(52) Zhukovsky, S. V.; Kidwai, O.; Sipe, J. E. Physical Nature of Volume Plasmon Polaritons in Hyperbolic Metamaterials. *Opt. Express* **2013**, *21* (12), 14982. https://doi.org/10.1364/oe.21.014982.

(53) Zhukovsky, S. V.; Andryieuski, A.; Sipe, J. E.; Lavrinenko, A. V. From Surface to Volume Plasmons in Hyperbolic Metamaterials: General Existence Conditions for Bulk High- k Waves in Metal-Dielectric and Graphene-Dielectric Multilayers. *Phys. Rev. B - Condens. Matter Mater. Phys.* **2014**, *90* (15), 155429. https://doi.org/10.1103/PhysRevB.90.155429.

(54) Kidwai, O.; Zhukovsky, S. V.; Sipe, J. E. Effective-Medium Approach to Planar Multilayer Hyperbolic Metamaterials: Strengths and Limitations. *Phys. Rev. A - At. Mol. Opt. Phys.* **2012**, *85* (5), 053842. https://doi.org/10.1103/PhysRevA.85.053842.

(55) Poddubny, A.; Iorsh, I.; Belov, P.; Kivshar, Y. Hyperbolic Metamaterials. *Nat. Photonics* **2013**, *7* (12), 948–957. https://doi.org/10.1038/nphoton.2013.243.

(56) Tschikin, M.; Biehs, S. A.; Messina, R.; Ben-Abdallah, P. On the Limits of the Effective Description of Hyperbolic Materials in the Presence of Surface Waves. *J. Opt. (United Kingdom)* **2013**, *15* (10), 105101. https://doi.org/10.1088/2040-8978/15/10/105101.

(57) Brzobohatý, O.; Čižmár, T.; Karásek, V.; Šiler, M.; Dholakia, K.; Zemánek, P. Experimental and Theoretical Determination of Optical Binding Forces. *Opt. Express* **2010**, *18* (24), 25389. https://doi.org/10.1364/oe.18.025389.

(58) Johnson, P. B.; Christy, R. W. Optical Constants of the Noble Metals. *Phys. Rev. B* **1972**, *6* (12), 4370–4379. https://doi.org/10.1103/PhysRevB.6.4370.

(59) Rodríguez-de Marcos, L. V.; Larruquert, J. I.; Méndez, J. A.; Aznárez, J. A. Self-Consistent Optical Constants of SiO_2 and Ta_2O_5 Films. *Opt. Mater. Express* **2016**, *6* (11), 3622. https://doi.org/10.1364/ome.6.003622.

**SUPPLEMENTARY**

GREEN'S FUNCTION FORMALISM

The two particles' interaction near the substrate can be described by three components of the Green's function. The Green's function $\hat{G}$ from Eq. (2) can be divided into three parts:

$\hat{G}^{fs}(\vec{r}_1,\vec{r}_2)$, $\hat{G}^{fs}(\vec{r}_2,\vec{r}_1)$ are responsible for particles' interaction in a free space;

$\hat{G}^{subs}(\vec{r}_1,\vec{r}_2)$, $\hat{G}^{subs}(\vec{r}_2,\vec{r}_1)$ describe particles' interaction via the substrate,

$\hat{G}^{si}(\vec{r}_1,\vec{r}_1)$, $\hat{G}^{si}(\vec{r}_2,\vec{r}_2)$ describe self-induced field, impact of the single particles' interaction with the substrate.

Thus, the expressions for these functions have the form [1]:

$$\hat{G}^{fs}(\vec{r}_1,\vec{r}_2) = \frac{\exp(ik_0 R)}{4\pi R}\left[\left(1+\frac{ik_0 R - 1}{(k_0 R)^2}\right)\hat{\mathbf{I}} + \frac{3-3ik_0 R - (k_0 R)^2}{(k_0 R)^2}\frac{\vec{R}\otimes\vec{R}}{R^2}\right], \quad (S1)$$

where $R$ is the length of a vector $\vec{R} = \vec{r}_1 - \vec{r}_2$, $\hat{\mathbf{I}}$ is unitary dyad.

$$\hat{G}^{si}(\vec{r}_1,\vec{r}_1) = \frac{i}{8\pi k_0^2}\int_0^\infty \frac{k_x}{k_{z1}}\begin{bmatrix} k_0^2 r^s - k_{z1}^2 r^p & 0 & 0 \\ 0 & k_0^2 r^s - k_{z1}^2 r^p & 0 \\ 0 & 0 & 2k_x^2 r^p \end{bmatrix} e^{2ik_{z1}z_0} dk_x. \quad (S2)$$

To simplify the form of the Green's function for the particles' interaction near the substrate one has to use normalized values $s = k_x/k_0$, $s_{z1} = k_{z1}/k_0$, $\tilde{z} = zk_0$ and transform the function to the cylindrical coordinates. Then, Bessel functions can be used.

$$\hat{G}^{subs}(\vec{r}_1,\vec{r}_2) = \frac{ik_0}{8\pi}\int_0^\infty \begin{pmatrix} m_{xx} & m_{xy} & m_{xz} \\ m_{yx} & m_{yy} & m_{yz} \\ m_{zx} & m_{zy} & m_{zz} \end{pmatrix} e^{is_{z1}\tilde{z}} ds \quad (S3)$$

$$m_{xx} = \frac{s}{s_{z1}}r^s f(s,\rho,\varphi) - ss_{z1}r^p g(s,\rho,\varphi),$$

$$m_{yy} = \frac{s}{s_{z1}}r^s g(s,\rho,\varphi) - ss_{z1}r^p f(s,\rho,\varphi),$$

$$m_{zz} = 2\pi J_0(s,\rho)r^p \frac{s^3}{s_{z1}},$$

$$m_{xy} = m_{yx} = \frac{r^s + r^p s_{z1}^2}{ss_{z1}} h(s,\rho,\varphi),$$

$$m_{xz} = -m_{zx} = -sr^p t(s,\rho,\varphi),$$

$$m_{yz} = -m_{zy} = -sr^p w(s,\rho,\varphi),$$

where the functions can be expressed via the Bessel ones:

$$f(s,\rho,\varphi) = 2\pi\left(\sin^2(\varphi)J_0(s\rho) + \frac{J_1(s\rho)}{s\rho}\cos(2\varphi)\right),$$

$$g(s,\rho,\varphi) = 2\pi\left(\cos^2(\varphi)J_0(s\rho) - \frac{J_1(s\rho)}{s\rho}\cos(2\varphi)\right),$$

$$h(s,\rho,\varphi) = \pi s^2 J_2(s\rho)\sin(2\varphi),$$

$$t(s,\rho,\varphi) = 2\pi i s J_1(s\rho)\cos(\varphi),$$

$$w(s,\rho,\varphi) = 2\pi i s J_1(s\rho)\sin(\varphi).$$

$J_n(z)$ is the first kind Bessel function of the order $n$.

EXPRESSIONS FOR THE EFFECTIVE FIELDS

To calculate an optical force acting on a particle one should previously obtain the value of the field $\vec{E}$ at a particle position. As it was mentioned in Eq. (2) of the main paper, this problem is self-consistent [1–3]. After substituting three kinds of the Green's function we get:

$$\vec{E}(\vec{r}_1) = \vec{E}_{inc}(\vec{r}_1) + \alpha_1 \frac{k_0^2}{\varepsilon_0} \hat{G}^{si}(\vec{r}_1,\vec{r}_1)\vec{E}(\vec{r}_1) + \alpha_2 \frac{k_0^2}{\varepsilon_0}\left[\hat{G}^{fs}(\vec{r}_1,\vec{r}_2) + \vec{G}^{subs}(\vec{r}_1,\vec{r}_2)\right]\vec{E}(\vec{r}_2); \qquad (S4)$$

$$\vec{E}(\vec{r}_2) = \vec{E}_{inc}(\vec{r}_2) + \alpha_1 \frac{k_0^2}{\varepsilon_0}\left[\hat{G}^{fs}(\vec{r}_2,\vec{r}_1) + \hat{G}^{subs}(\vec{r}_2,\vec{r}_1)\right]\vec{E}(\vec{r}_1) + \alpha_2 \frac{k_0^2}{\varepsilon_0} \hat{G}^{si}(\vec{r}_2,\vec{r}_2)\vec{E}(\vec{r}_2); \qquad (S5)$$

where $\alpha_1$, $\alpha_2$ are polarizabilities of the particles. As long as we use identical particles, these values are equal. These equations should be viewed as a system. After substituting Eq. (S5) into Eq. (S4) or vice versa we get the final form of the expression:

$$\vec{E}(\vec{r}_1) = \left[\hat{g}(\vec{r}_1,\vec{r}_1) - \hat{G}(\vec{r}_1,\vec{r}_2)\hat{g}(\vec{r}_2,\vec{r}_2)^{-1}\hat{G}(\vec{r}_2,\vec{r}_1)\right]^{-1} \left(\vec{E}_{inc}(\vec{r}_1) + \hat{G}(\vec{r}_1,\vec{r}_2)\hat{g}(\vec{r}_2,\vec{r}_2)^{-1}\vec{E}_{inc}(\vec{r}_2)\right) \quad (S6)$$

$$\vec{E}(\vec{r}_2) = \left[\hat{g}(\vec{r}_2,\vec{r}_2) - \hat{G}(\vec{r}_2,\vec{r}_1)\hat{g}(\vec{r}_1,\vec{r}_1)^{-1}\hat{G}(\vec{r}_1,\vec{r}_2)\right]^{-1} \left(\vec{E}_{inc}(\vec{r}_2) + \hat{G}(\vec{r}_2,\vec{r}_1)\hat{g}(\vec{r}_1,\vec{r}_1)^{-1}\vec{E}_{inc}(\vec{r}_1)\right); \qquad (S7)$$

where

$$\hat{g}(\vec{r}_1,\vec{r}_1) = \hat{\mathbf{I}} - \alpha_1 \frac{k_0^2}{\varepsilon_0} \hat{G}^{si}(\vec{r}_1,\vec{r}_1);$$

$$\hat{g}(\vec{r}_2,\vec{r}_2) = \hat{\mathbf{I}} - \alpha_2 \frac{k_0^2}{\varepsilon_0} \vec{G}^{si}(\vec{r}_2,\vec{r}_2);$$

$$\hat{G}(\vec{r}_1,\vec{r}_2) = \alpha_2 \frac{k_0^2}{\varepsilon_0}\left[\hat{G}^{fs}(\vec{r}_1,\vec{r}_2) + \vec{G}^{subs}(\vec{r}_1,\vec{r}_2)\right];$$

$$\hat{G}(\vec{r}_2,\vec{r}_1) = \alpha_1 \frac{k_0^2}{\varepsilon_0}\left[\hat{G}^{fs}(\vec{r}_2,\vec{r}_1) + \hat{G}^{subs}(\vec{r}_2,\vec{r}_1)\right].$$

THE IMPACT OF THE TOPMOST LAYER

One of the possible realization of hyperbolic metamaterial is metal-dielectric multilayered structure. We consider a multilayered structure with Ag [4] and $Ta_2O_5$ [5] layers, filling factor $f = 1:7.5$ and thickness of the slab is $d = \lambda_0/8$. The reflection from the structure of this type can be calculated with the help of transfer matrix formalism [6]. One can compare the optical binding force near the structure with 10 periods with metal (Figure 1S, red line) or dielectric (Figure 1S, blue line) layer on top. One can see that for the silver layer on top of the structure optical binding force is greater than for the case of the dielectric layer. The topmost metallic layer provides better coupling between the surface and volumetric modes, which leads to the hyperbolic mode contribution increase.

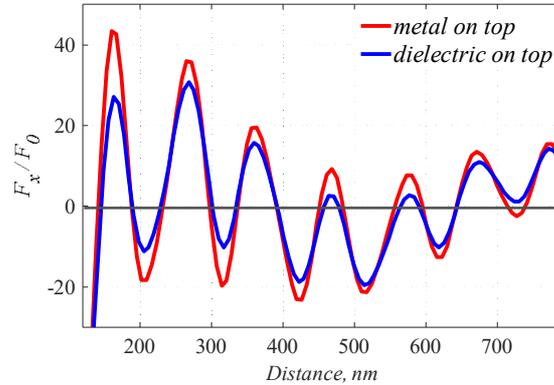

**Figure 1S**. Optical binding force near hyperbolic metamaterial slab. Red and blue lines correspond to the multilayered structure with metal or dielectric top layer correspondingly. The Figure represents the structure with $d = \lambda_0/8$. Incident light wavelength is 920 nm, $f = 1:7.5$.

## SPATIAL DISTRIBUTION OF THE ELECTROMAGNETIC FIELD ABOVE THE THIN SLAB

The specific character of optical binding force near anisotropic slab is defined by electrical field distribution. Here we consider *x*-component of this field near the slab with $d = \lambda_0 / 8$, and parameters corresponding to *A* and *C* ($\lambda_0 = 920$ nm) from the main paper (Figure 5). Single particle near the substrate is illuminated by a plane wave, and field distributions in two planes (top ((a), (b)) and front view ((c), (d))) are shown. In Figure 2S high-intensity regions of volumetric modes are clearly seen in both views. The high-intensity regions are aligned along incident light polarization. Figures 2S (a), (c) correspond to *A* while (b), (d) correspond to *C*. The result is obtained in COMSOL Multiphysics package.

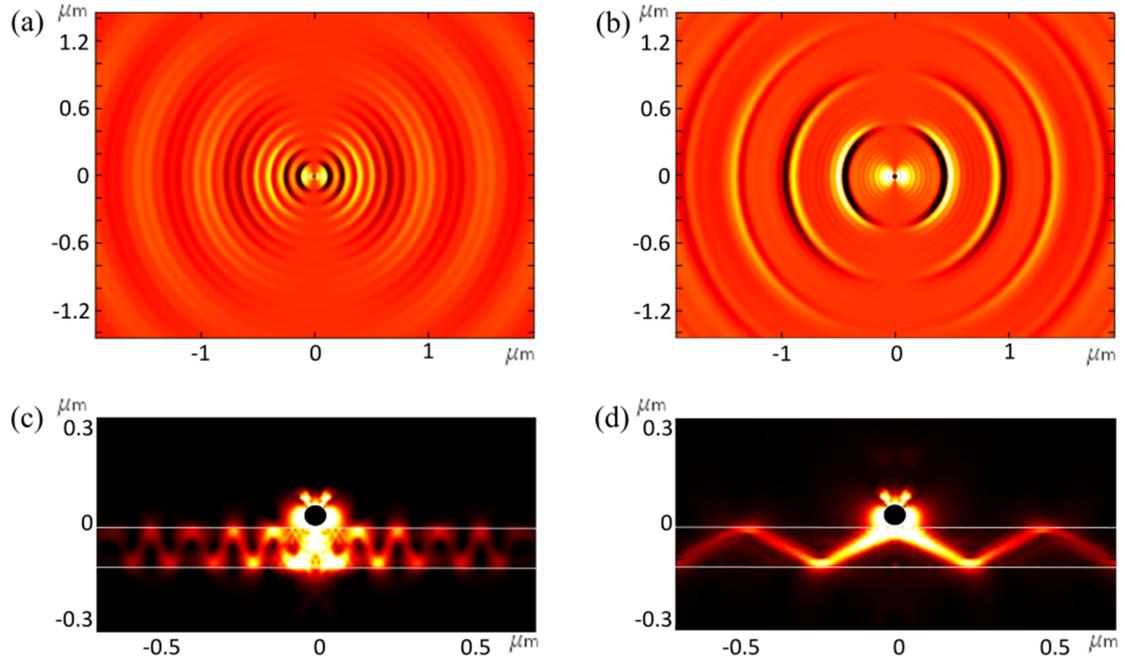

**Figure 2S.** Field distribution for a single particle above the metamaterial layer. (a), (b) *XY* plane, top view; (c), (d) *XZ* plane, front view. Field intensity is given in arbitrary units.

**REFERENCES**

(1) Novotny, L.; Hecht, B. *Principles of Nano-Optics*; 2009; Vol. 9781107005. https://doi.org/10.1017/CBO9780511794193.

(2) Kostina, N.; Petrov, M.; Ivinskaya, A.; Sukhov, S.; Bogdanov, A.; Toftul, I.; Nieto-Vesperinas, M.; Ginzburg, P.; Shalin, A. Optical Binding via Surface Plasmon Polariton Interference. *Phys. Rev. B* **2019**, *99* (12), 125416. https://doi.org/10.1103/PhysRevB.99.125416.

(3) Petrov, M. I.; Sukhov, S. V.; Bogdanov, A. A.; Shalin, A. S.; Dogariu, A. Surface Plasmon Polariton Assisted Optical Pulling Force. *Laser Photonics Rev.* **2016**, *10* (1), 116–122. https://doi.org/10.1002/lpor.201500173.

(4) Johnson, P. B.; Christy, R. W. Optical Constants of the Noble Metals. *Phys. Rev. B* **1972**, *6* (12), 4370–4379. https://doi.org/10.1103/PhysRevB.6.4370.

(5) Rodríguez-de Marcos, L. V.; Larruquert, J. I.; Méndez, J. A.; Aznárez, J. A. Self-Consistent Optical Constants of SiO_2 and Ta_2O_5 Films. *Opt. Mater. Express* **2016**, *6* (11), 3622. https://doi.org/10.1364/ome.6.003622.

(6) Coldren, L. A. *Diode Lasers and Photonic Integrated Circuits*; 1997; Vol. 36. https://doi.org/10.1117/1.601191.